\documentclass[prl,aps,floatfix,twocolumn,superscriptaddress,longbibliography,10pt]{revtex4-2}

\usepackage{amsmath,amssymb}
\usepackage{graphicx}
\usepackage[dvipsnames]{xcolor}
\usepackage[]{hyperref}
\usepackage{amsthm}
\usepackage{enumerate}
\usepackage{setspace}
\usepackage[export]{adjustbox}
\usepackage{dsfont}
\usepackage{braket}
\hypersetup{
    pdfstartview={FitH}, 
    colorlinks=true, 
    linkcolor=blue, 
    citecolor=blue, 
    filecolor=magenta,
    urlcolor=blue
}

\usepackage[capitalize]{cleveref}
\usepackage{bm}
\usepackage{mathtools}
\usepackage{url}
\usepackage{enumitem}

\usepackage{verbatim}
\usepackage{MnSymbol}
\usepackage{physics}
\usepackage{stmaryrd}

\begin{document}

\title{Neural Decoders for Universal Quantum Algorithms}

\author{J. Pablo Bonilla Ataides}
\thanks{These authors contributed equally.}
\affiliation{Department of Physics, Harvard University, Cambridge, MA 02138, USA}
\email{jbonillaataides@g.harvard.edu; andigu@g.harvard.edu; syelin@g.harvard.edu; lukin@physics.harvard.edu}

\author{Andi Gu}
\thanks{These authors contributed equally.}
\affiliation{Department of Physics, Harvard University, Cambridge, MA 02138, USA}

\author{Susanne F. Yelin}
\affiliation{Department of Physics, Harvard University, Cambridge, MA 02138, USA}

\author{Mikhail D. Lukin}
\affiliation{Department of Physics, Harvard University, Cambridge, MA 02138, USA}

\thanks{$^*$These authors contributed equally.}

\begin{abstract}
Fault-tolerant quantum computing demands decoders that are fast, accurate, and adaptable to circuit structure and realistic noise. While machine learning (ML) decoders have demonstrated impressive performance for quantum memory, their use in algorithmic decoding---where logical gates create complex error correlations---remains limited. We introduce a modular attention-based neural decoder that learns gate-induced correlations and generalizes from training on random circuits to unseen multi-qubit algorithmic workloads. Our decoders achieve fast inference and logical error rates comparable to most-likely-error (MLE) decoders across varied circuit depths and qubit counts. Addressing realistic noise, we incorporate loss-resolving readout, yielding substantial gains when qubit loss is present. We further show that by tailoring the decoder to the structure of the algorithm and decoding only the relevant observables, we can simplify the decoder design without sacrificing accuracy.
We validate our framework on multiple error correction codes---including surface codes and 2D color codes---and demonstrate state-of-the-art performance under circuit-level noise. Finally, we show that the use of attention offers interpretability by identifying the most relevant correlations being tracked by the decoder. Enabling experimental validation of deep-circuit fault-tolerant algorithms and architectures (Bluvstein \textit{et al.}, arXiv:2506.20661, 2025), these results establish neural decoders as practical, versatile, and high-performance tools for quantum computing.
\end{abstract}

\maketitle

Fault-tolerant quantum computation requires quantum error correction (QEC) to suppress the effects of noise below the threshold needed for practical algorithms~\cite{gottesman1997stabilizer,dennis2002topological,kitaev2003fault}. In stabilizer codes, continuous syndrome measurements extract error information, which a decoder must interpret to compute appropriate recovery operations. As quantum computing advances from proof-of-principle demonstrations to practical applications, the demands on decoders are becoming increasingly stringent: they must be fast enough for real-time operation, accurate enough to maintain sub-threshold performance, and flexible enough to handle diverse algorithmic structures and realistic noise models.

Machine learning (ML) decoders have emerged as promising solutions, offering fast inference times and competitive performance compared to conventional alternatives~\cite{krastanov2017deep,varsamopoulos2017decoding,torlai2017neural,baireuther2018machine,chamberland2018deep,baireuther2019neural,maskara2019advantages,andreasson2019quantum,sweke2020reinforcement,varsamopoulos2020decoding,domingo2020reinforcement,ni2020neural,fitzek2020deep,wagner2020symmetries,meinerz2022scalable,wang2023transformer,gicev2023ascalable,chamberland2023techniques,bausch2024learning,lange2025data,varbanov2025neural}. These architectures combine hardware efficiency with neural network adaptability, being deployable on FPGAs or ASICs for real-time decoding~\cite{overwater2022neural,boutros2024field,coelho2021automatic}; they further enable training directly on experimental data and  fine-tuning simulated models~\cite{lange2025data,bausch2024learning,google2024quantum,bluvstein2025architectural}.

Despite these advantages, the vast majority of prior work has focused on the conceptually simpler problem of \emph{quantum memory} --- maintaining a logical state over time while performing active error correction. In this setting, qubits remain idle between syndrome measurements, and errors accumulate independently on stationary physical qubits. %
The transition to \emph{algorithmic decoding} --- correcting errors during the execution of logical quantum algorithms --- presents fundamentally new challenges. During algorithm execution, logical gates induce correlations between errors on different qubits, creating structured error patterns that propagate through the circuit. These gate-dependent correlations must be captured by the decoder to prevent substantial performance degradation~\cite{cain2024correlated,cain2025fast}. Moreover, practical quantum algorithms must operate under realistic noise models that include processes such as qubit loss and leakage. These features are either largely absent or difficult to incorporate into existing decoder frameworks.

Recent experimental progress has made these challenges increasingly urgent. Demonstrations of logical quantum algorithms on various platforms~\cite{egan2021fault,ryananderson2021realization,zhao2022realization,krinner2022realizing,google2023suppressing,bluvstein2023logical,sivak2023real,gupta2024encoding,google2024quantum} have highlighted the need for improved decoding capabilities in order to meet the requirements for fault-tolerant computation. While specialized decoders exist for quantum memory~\cite{fowler2013optimal,bravyi2014efficient,huang2020fault,delfosse2021almost,higgot2023improved,sundaresan2023demonstrating,higgott2025sparse,barber2025areal} and there has been progress toward decoding algorithmic circuits~\cite{cain2024correlated,zhou2025low,cain2025fast,serraperalta2025decoding,zhou2025learning,turner2025scalable}, no existing framework provides the combination of speed, accuracy, and flexibility needed for practical algorithmic decoding.

\begin{figure}[h!]
    \centering
    \includegraphics[width=\columnwidth]{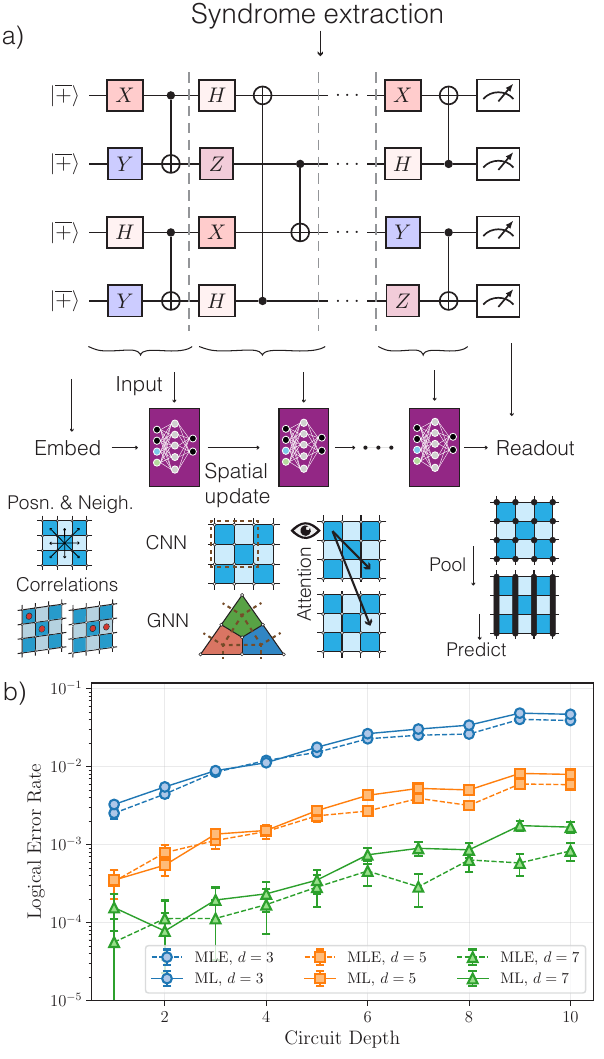}
    \caption{\textbf{Neural decoder architecture and performance.} (a) Schematic of the neural decoder, which processes quantum error correction circuits to estimate logical errors on each qubit. The input comprises stabilizer measurement circuits with embedded gate and syndrome information. Each stabilizer is assigned a vector representation that evolves via recurrent and spatial updates through the circuit. Spatial processing is implemented using convolutional layers for surface codes and graph neural networks for color codes, with an attention block continuously tracking correlations between logical qubits. The final readout aggregates stabilizer embeddings through pooling to predict logical error probabilities. (b) Decoder performance as a function of circuit depth and code distance, comparing the machine-learned decoder (ML) to a most-likely-error (MLE) decoder. Data are presented as mean values plus or minus standard deviations.}
    \label{fig:neural-decoder}
\end{figure}

In this Article, we develop neural decoders specifically tailored for universal quantum algorithms operating under realistic noise conditions. Our approach addresses algorithmic decoding challenges by using an attention-based, modular neural architecture---where each logical gate is a distinct trainable module---that jointly tracks error correlations across all logical qubits, enabling fast inference with accuracy comparable to an exact MLE decoder, whose inference cost grows exponentially with system size. Because the architecture learns to model error propagation through individual gates, a single trained model can be automatically applied to new algorithmic circuits, without the need for retraining whenever the logical circuit changes. We further develop qubit loss-aware decoders that exploit loss-resolving readout for improved performance under qubit loss, and combine decoder-level robustness with algorithmic strategies like logical teleportation to prevent loss accumulation. By leveraging recent insights that decoding can be greatly simplified by targeting only the relevant observables of the circuit~\cite{cain2025fast} and by exploiting constrained error propagation, we design specialized, efficient decoders for deep cluster-state circuits, enabling high-performance error correction in measurement-based quantum computing architectures. Importantly, we demonstrate a degree of interpretability in our attention-based architecture: by probing the attention weights, we find the model learns to focus on physically meaningful correlations, consistent  with the expected propagation of errors through entangling gates in the circuit.
Collectively, these advances establish neural decoders as a robust, accurate, and practical foundation for fault-tolerant quantum algorithms under realistic circuit- and noise-level constraints.

Importantly, the neural decoder methods introduced and analyzed here have already been deployed in recent experimental tests of deep-circuit fault-tolerant algorithms and architectures, demonstrating superior performance on both quantum memory tasks with surface codes and deep-circuit algorithms involving complex encodings~\cite{bluvstein2025architectural}.
The present work systematically generalizes and extends these methods, uncovering the architectural features underpinning their efficient performance and demonstrating that these decoders can be scaled up and flexibly applied to a broad range of quantum computing tasks.

\vspace{-5mm}
\section{Neural Decoder Architecture}
\vspace{-4mm}
Our neural decoder architecture (Fig.~\ref{fig:neural-decoder}(a)) processes error-corrected logical circuits to predict logical errors on individual logical qubits. The decoder takes as input the complete syndrome history from a logical circuit execution and outputs, for each logical qubit, the marginal probability that the qubit has experienced a logical error. The architecture is built around three core components: (1) a structured input representation that captures both syndrome measurements and circuit structure, (2) an internal per-stabilizer embedding that evolves through the circuit to capture error correlations, and (3) a spatial readout mechanism that aggregates stabilizer information to produce logical error predictions. The key insight underlying our design is that quantum error correction fundamentally involves tracking how errors propagate through both space and time. Spatially, errors spread between nearby qubits through gate operations and stabilizer measurements. Temporally, errors accumulate and correlate as the circuit progresses. Our architecture explicitly models both dimensions: the internal representation captures the temporal evolution of error information, while spatial processing operations capture how this information should be shared between nearby stabilizers. It consists of the following specific elements. 
\subsection{Input and Output Format}
\vspace{-4mm}
The decoder takes as input a complete description of a logical quantum circuit execution:
\begin{itemize}
\item \textbf{Syndrome measurements}: For each stabilizer measurement round, we provide the measurement outcomes for all stabilizers across all logical qubits. This includes both X-type and Z-type stabilizer measurements, along with any flag measurements used in the error correction protocol.
\item \textbf{Circuit structure}: We provide the sequence of logical operations (single-qubit gates, two-qubit gates) applied to each logical qubit, along with connectivity information indicating which pairs of logical qubits interact during two-qubit operations.
\item \textbf{Code parameters}: The code distance and type (surface code or color code) for each logical qubit.
\item \textbf{Noise parameters}: A compact representation of the noise model used in the simulation.
\end{itemize}
The decoder outputs the marginal probability that each logical qubit indicating the likelihood that it has experienced a logical error. Specifically, for circuits ending with computational basis measurements, this represents the probability of a logical X error (bit flip). For circuits ending with X-basis measurements, this represents the probability of a logical Z error (phase flip).
\subsection{Internal Representation}
\vspace{-4mm}
The core of our architecture is a learned \textit{per-stabilizer embedding} that tracks error-relevant information for each stabilizer throughout the circuit execution. Each stabilizer --- whether X-type or Z-type, and whether associated with syndrome extraction or final measurements --- maintains a fixed-dimensional vector representation that evolves as the circuit progresses.
These embeddings are designed to capture several types of information:
\begin{itemize}
\item \textbf{Local syndrome history}: The pattern of measurement outcomes for this stabilizer and its neighbors over time.
\item \textbf{Geometric metadata}: The stabilizer's position within the error correction code, as well its connectivity to neighboring stabilizers.
\item \textbf{Temporal correlations}: How errors affecting this stabilizer may have propagated from earlier time steps.
\item \textbf{Inter-qubit correlations}: For multi-qubit circuits, how this stabilizer's errors may be correlated with stabilizers on other logical qubits due to entangling operations.
\end{itemize}
The stabilizer embeddings are updated through two complementary mechanisms. First, a recurrent update incorporates new syndrome information as each layer of the circuit is processed, allowing the embedding to accumulate information over time. Second, spatial processing operations allow stabilizers to share information with their neighbors, capturing how errors propagate through the code's geometric structure.
For surface codes, spatial processing is implemented using two-dimensional convolutions that reflect the grid structure of the code. For color codes, we use graph neural networks~\cite{schlichtkrull2017modeling} that operate on the more complex connectivity pattern of the hexagonal lattice. In both cases, the spatial operations are designed to model how detection events at one stabilizer should influence the error estimates at nearby stabilizers, and these correlations are tracked internally through the use of attention.

\subsection{Spatial Readout and Prediction}
\vspace{-4mm}
To produce logical error predictions, the decoder must aggregate information from all stabilizers associated with each logical qubit. This aggregation respects the structure of the logical operators: for computational basis measurements, we focus on information relevant to X-type logical errors, while for X-basis measurements, we focus on Z-type logical errors. The readout process first converts the per-stabilizer embeddings into per-data-qubit representations through spatial processing operations. For surface codes, this involves additional two-dimensional convolutions that map stabilizer information onto the data qubits they surround. For color codes, we use specialized graph convolutions that propagate information from stabilizer faces to the data qubits on their boundaries. Next, we perform \textit{logical observable pooling}: we aggregate the data qubit representations along the support of the relevant logical operator. For example, when predicting X logical errors in a surface code, we pool information from data qubits along a column of the grid (the support of the Z logical operator, which anticommutes with X logical errors). For the surface code, we pool along each of the appropriate row/column representatives of the logical operator (see Fig.~\ref{fig:neural-decoder}(a)), whereas for the color code pooling is conducted along a fixed logical representative. This pooling operation produces a single vector representation for each logical qubit. Finally, a fully connected network processes this aggregated representation to produce the final logical error probability. The network is trained to output well-calibrated probabilities that can be used both for error correction decisions and for confidence estimation in downstream applications~\cite{bausch2024learning,bluvstein2025architectural}.

\subsection{Training Methodology}
\vspace{-4mm}
While the decoder architecture is designed for end-to-end circuit processing, we employ several training techniques to improve data efficiency and generalization.

\textbf{Intermediate supervision}: Rather than only providing supervision at the end of each circuit, we use intermediate measurements to create additional training examples. For a circuit with $n$ layers, we simulate what would happen if transversal measurements were performed after each layer $i \in \qty{1,2,\ldots,n}$, creating $n$ distinct training examples from a single circuit. This trick, also used in Ref.~\cite{bausch2024learning}, dramatically increases the effective size of our training dataset.

\textbf{On-the-fly data generation}: We generate training circuits randomly during training, randomly sampling the code distance, number of logical qubits, circuit depth, and gate sequence. This ensures the model sees a diverse range of data during training.

\textbf{Permutation invariance}: To ensure the model can handle arbitrary numbers of logical qubits, we assign each logical qubit a ``fingerprint'' embedding, which is randomly generated for each training shot, that is used throughout the circuit processing. This prevents the model from learning position-dependent behaviors that would not generalize to different numbers of qubits, and instead forces the model to learn how correlations spread between logical qubits as a function of the underlying logical circuit. Further details are provided in the Methods.

\subsection{Inference}
\vspace{-4mm}

After the neural decoders have been trained, they can be deployed for inference on multi-qubit logical algorithms. In \cref{fig:neural-decoder}(b), we show the logical error rate as a function of circuit depth for the surface code with $d=3,5,7$, comparing our neural decoder against an exact implementation of a decoder that identifies the most-likely-error (MLE) compatible with the observed syndrome. The MLE decoder achieves state-of-the-art logical error rates, but its computational cost increases exponentially with system size~\cite{cain2024correlated}. Remarkably, the trained neural decoder attains logical error rates approaching those of the MLE decoder. This highlights the flexibility and accuracy of our architecture: a single trained model generalizes across all code distances, depths, and circuit instances while maintaining state-of-the-art accuracies. The speed of inference is summarized in \cref{fig:timing}(a); decoding shot-by-shot yields a time per gate of approximately $27$ ms, which can be reduced to around $100~\mu$s by batching syndromes and utilizing GPU parallelism. Additional details of these simulations and timing benchmarks are provided in the Methods.

\section{Loss Decoding}
\begin{figure}[h!]
    \centering
    \includegraphics[width=\columnwidth]{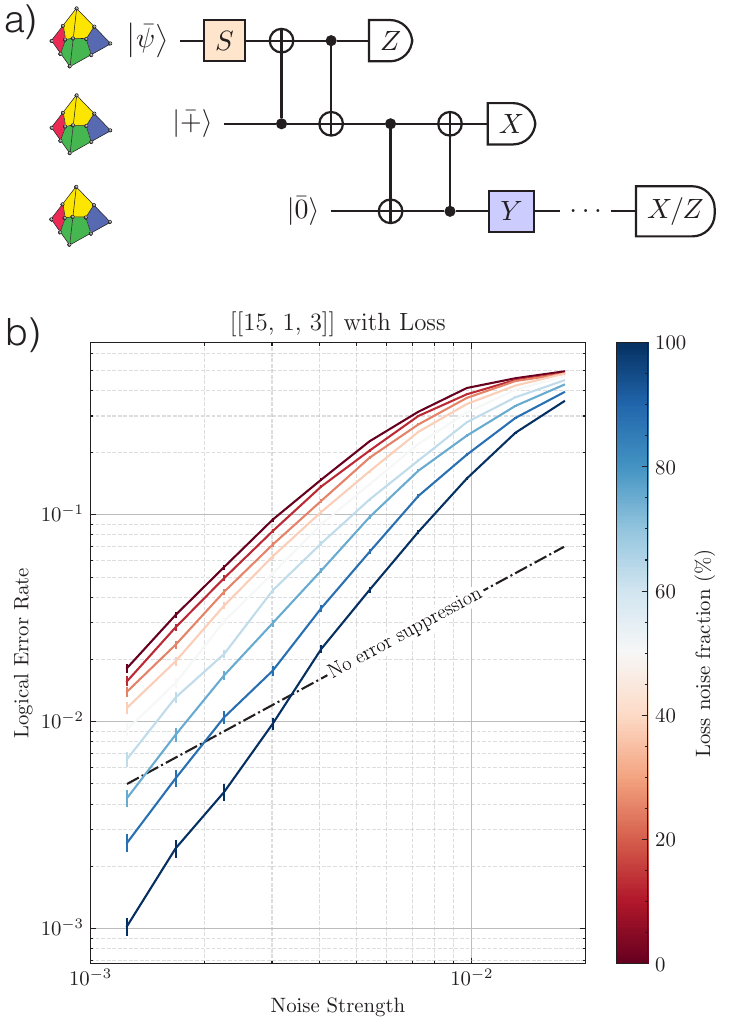}
    \caption{\textbf{Loss-aware decoding with neural decoders.} (a) Circuit schematic for a logical qubit encoded in a $[[15,1,3]]$ Reed-Muller code, undergoing alternating layers of random transversal Clifford gates and logical teleportation, subject to qubit loss events. Logical teleportation steps integrate loss-resolving readout (LRR), enabling the identification and correction of lost qubits. (b) Logical error rate as a function of noise strength for different loss noise fractions (a loss noise fraction of $100\%$ indicates all noise in the system is comprised of loss, while a loss noise fraction of $0\%$ indicates all noise is comprised of Pauli noise). As the loss fraction increases, the loss-aware decoder shows greater benefit, with steeper declines in logical error rate and crossings below the no error suppression line occurring at higher noise strengths.}
    \label{fig:teleported-memory}
\end{figure}

\vspace{-4mm}
In this section, we show that explicitly incorporating loss information into the decoder significantly enhances error correction performance, highlighting the flexibility of neural decoders to handle realistic quantum noise processes. As an example, inspired by the recent experiment of Ref.~\cite{bluvstein2025architectural}, we decode circuits in which a logical qubit, encoded with a $[[15,1,3]]$ Reed-Muller (RM) code, undergoes alternating rounds of random transversal Clifford gates and logical teleportation, as illustrated in \cref{fig:teleported-memory}(a). We decode circuits in which a logical qubit, encoded with a $[[15,1,3]]$ Reed-Muller (RM) code, undergoes alternating rounds of random transversal Clifford gates and logical teleportation, as illustrated in \cref{fig:teleported-memory}(a). Logical teleportation is implemented via a transversal measurement followed by a classically conditioned logical Clifford correction (or, equivalently, by an added transversal CNOT before measurement), and serves as a key algorithmic primitive to prevent the accumulation of loss~\cite{bluvstein2025architectural}. At each teleportation step, lost qubits are detected using Loss-Resolving Readout (LRR), ensuring that loss does not accrue over time. The transversal measurements further enable code checks to be reconstructed and corrections determined, removing the need for repeated rounds of active error correction. 

\begin{figure*}[ht]
    \centering
    \includegraphics[width=\textwidth]{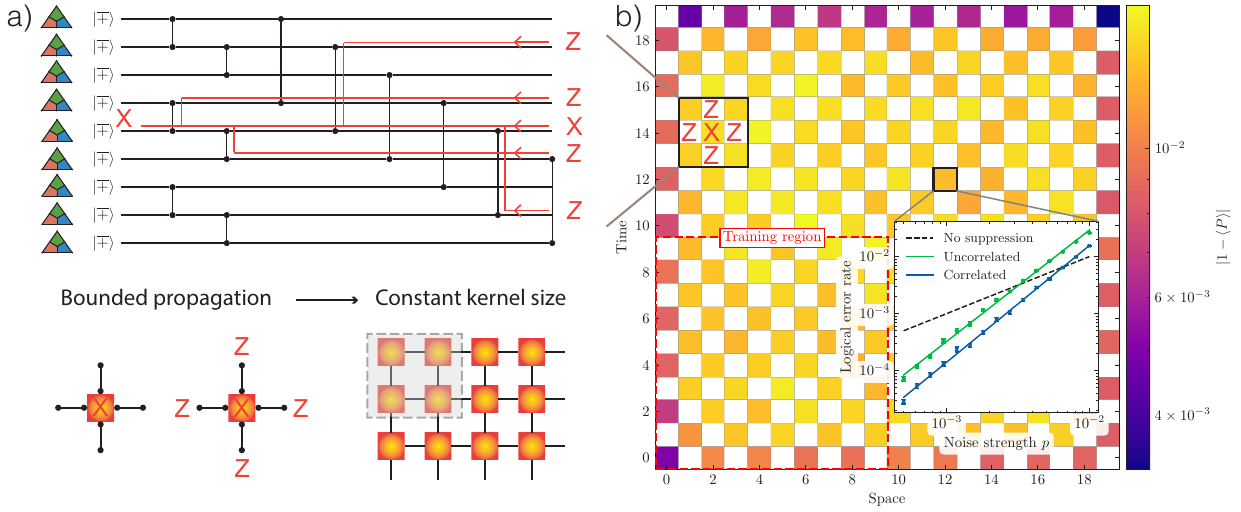}
    \caption{\textbf{Deep logical cluster circuits.} (a) Schematic of 2D cluster-state circuits built from logical 7-qubit Steane codes. Error propagation is strictly local due to the commutation properties of the gates. (b) Decoder performance for cluster states of size $20\times20$ at noise strength $p=0.01$, after training only on clusters up to size $10\times10$. The decoder predicts the \emph{relevant observables} of the circuit---specifically, the cluster-state stabilizers, which are logical Pauli products comprising logical $\overline{X}$ on each qubit and logical $\overline{Z}$ on neighboring qubits.
    The results demonstrate generalization to larger systems and greater circuit depths. The inset compares a convolutional decoder exploiting local, observable-specific correlations (``correlated'') versus an architecture that does not (``uncorrelated''), highlighting the substantial performance gain from decoding only those observables and qubits with nontrivial short-range error correlations.}
    \label{fig:deep}
\end{figure*}

During training, we simulate measurement outcomes from circuits in which each loss event is approximated by the application of a fully depolarizing channel---effectively modeling the loss as a random Pauli error. This ``erasure'' approximation enables significantly faster data generation, and furthermore allows us to train the neural decoder using the same architecture described earlier, with minimal modification to the data format. Importantly, this approximation is well-justified in our setting: logical information in our circuits is frequently teleported, so there are few possible locations for a loss event to occur before detection and teleportation, limiting the impact of any unmodeled propagation of loss.

For inference, however, we simulate physical loss \emph{exactly}: lost qubits are removed from the circuit and any subsequent gates acting on them are omitted, so the deployed decoder is evaluated on data generated with a fully realistic noise model. We further assume access to loss-resolving readout (LRR), allowing the decoder to distinguish among measurement outcomes 0, 1, and L (lost)~\cite{baranes2025leveragingatomlosserrors,bluvstein2025architectural}. The neural network is equipped with a distinct embedding for each possible outcome, ensuring that the model can effectively utilize loss information to improve logical error rates.

Figure~\ref{fig:teleported-memory}(b) displays the logical error rate as a function of noise strength for different loss fractions. As the loss rate increases, the advantage of explicit loss-resolving readout becomes more apparent: the decoder achieves lower logical error rates and is able to maintain error suppression well below the ``no error suppression” regime, even at higher noise levels. This is consistent with both theoretical and experimental observations indicating that while undetected loss errors can severally limit the performance of FTQC, if detected and properly decoded they can result in increased error thresholds~\cite{wu2022erasure,ma2023high,scholl2023erasure,kubica2023erasure,baranes2025leveragingatomlosserrors,bluvstein2025architectural}. These results highlight the importance of accurately modeling loss in quantum circuits and demonstrate the capability of machine learning decoders to flexibly adapt to, and exploit, complex noise processes in realistic quantum hardware.

\section{Constraining attention}
\vspace{-4mm}
We now leverage recent insights into algorithmic decoding to further enhance our decoder for circuits of particular interest~\cite{cain2025fast}. Specifically, we demonstrate that by tailoring the decoder to focus on the relevant observables of a quantum circuit---and by constraining the search for correlations to only the qubits associated with those observables---the architecture can be significantly simplified. This reduction in complexity diminishes the need for computationally expensive all-to-all attention mechanisms, resulting in substantial improvements in decoding speed.

Again motivated by recent experiments~\cite{bluvstein2025architectural}, we focus on one- and two-dimensional cluster state circuits constructed from logical Steane codes~\cite{steane1996multiple}, where logical qubits are initialized in the $\ket{+}$ state and entangled by nearest-neighbor controlled-$Z$ (CZ) gates. In this setting, decoding can be simplified by restricting to \emph{reliable logical Pauli products}~\cite{cain2025fast}---in particular, the cluster state logical stabilizers, which are products of logical $\overline{X}$ on each qubit and logical $\overline{Z}$ on its neighbors.

Backpropagating these logical stabilizers through the circuit shows that each stabilizer terminates as a logical $\overline{X}$ operator on the corresponding input $\ket{\overline{+}}$ state (see Fig.~\ref{fig:deep} (a)). Thus, predicting the value of each cluster state stabilizer is equivalent to predicting the sign of the initial input state for that logical qubit. We set this sign of the initial input state as the objective function during training. To ensure balanced training, we randomize input states between $\ket{\overline{+}}$ and $\ket{\overline{-}}$.

The key feature enabling this simplification is the constrained propagation of errors: $Z$-type errors commute through the CZ gates, while $X$-type errors transform into $XZ$, with the additional $Z$ passing to the neighboring qubit. This mechanism ensures that error correlations remain strictly short-range in both space and time. Consequently, only the middle $\overline{X}$ qubit and its neighboring $\overline{Z}$ qubits are relevant for decoding each stabilizer observable.

Leveraging this locality, we replace the full attention-based architecture with a compact convolutional neural network (CNN) decoder with a kernel size of three, sufficient to capture all relevant correlations (Fig.~\ref{fig:deep} (a)). Decoder performance is assessed by decoding the  cluster state logical stabilizers, corresponding to logical $X$ on each qubit and logical $Z$ on its neighbors. Figure~\ref{fig:deep} (b) shows decoded stabilizer values for both the CNN architecture and, for comparison, an architecture lacking explicit access to local correlations. The CNN decoder, exploiting this structure, achieves superior performance. Furthermore, by measuring logical qubits in alternating $X$ and $Z$ bases, we recover half the logical stabilizers at any given time, with each decoded independently.

\begin{figure*}[ht]
    \centering
    \includegraphics[width=0.7\linewidth]{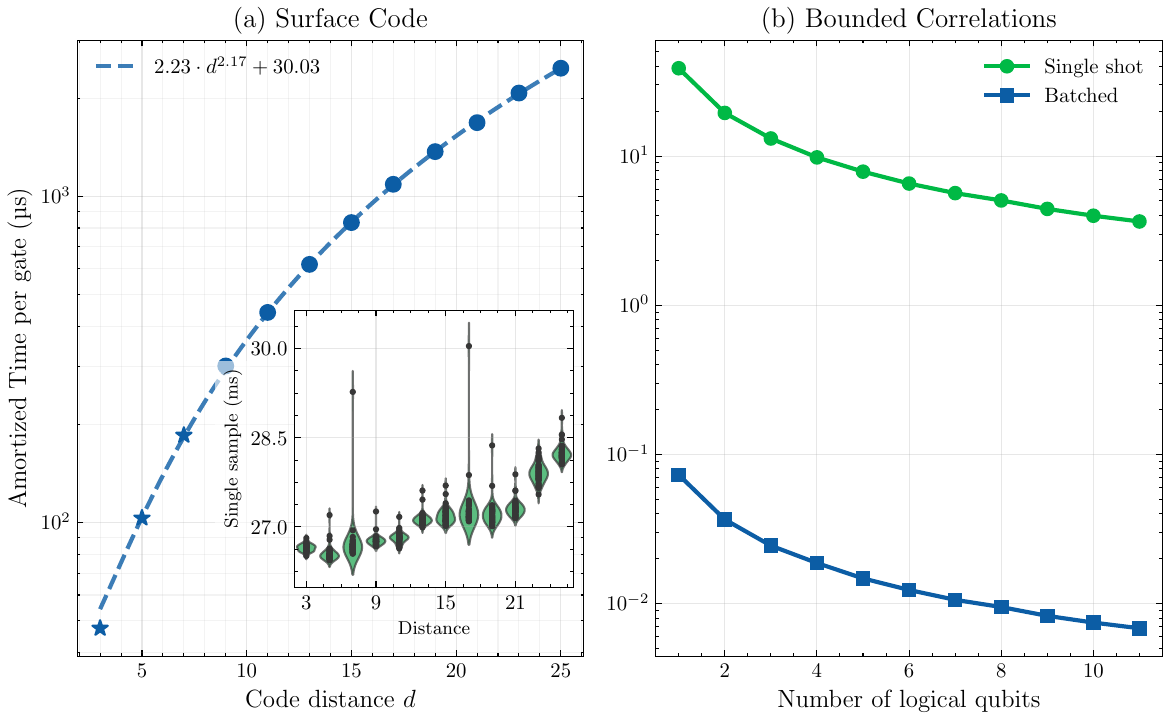}
    \caption{\textbf{Neural decoder inference time for random Clifford circuits.} The main panel in (a) shows the averaged (amortized) inference time per sample as a function of code distance, using batched evaluation on a NVIDIA H100 GPU. The stars indicate distances for which we have evaluated logical error rates (see \cref{fig:neural-decoder}). The inset displays single-sample inference times. For batched-inference, scaling is dominated by attention operations but is mitigated in practice by GPU optimizations, resulting in near-quadratic scaling with code distance. However, single-shot performance is almost independent of code distance, as a result of the high degree of parallelism on GPUs. On the other hand, in (b), we show the decoding time per qubit, per gate (as a function of the number of logical qubits), for the decoder architecture used to decode the cluster state circuit with Steane codes. Importantly, this architecture does not use attention, and employs only convolutions to model error correlations. Not only is this architecture significantly faster, the natural parallelism of the convolution mechanism means the total decoding time depends very weakly on the number of logical qubits (hence the decrease in amortized decoding time as a function of the number of logical qubits). For the `batched' timing on both plots, the batch size is 256.}
    \label{fig:timing}
\end{figure*}

In summary, by identifying the relevant reliable observables of the circuit and focusing decoding only on those qubits that exhibit nontrivial correlations---guided by the framework of fast correlated decoding~\cite{cain2025fast} --- we are able to construct compact, efficient decoders specifically tailored for deep cluster circuits. This targeted strategy enables high accuracy with significantly reduced model complexity and much faster inference than general-purpose architectures. For instance, \cref{fig:timing} demonstrates that by appropriately constraining attention, inference speed can increase by up to 3 orders of magnitude. Further details regarding inference time can be found in Methods. These findings emphasize that exploiting observable-specific correlations should be a central methodology when developing decoders for circuits with well-defined reliable observables.

\begin{figure}[h!]
\begin{center}
\includegraphics[width=\columnwidth]{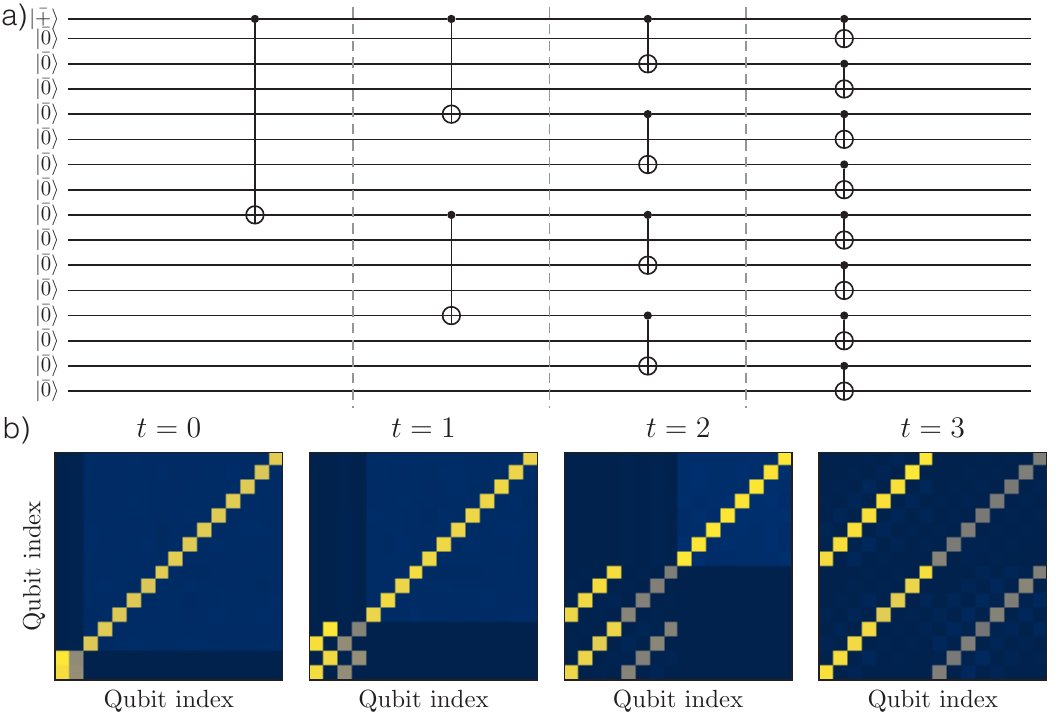}
\caption{{\bf Attention interpretability.} 
(a) Circuit diagram showing sequential layers of entangling gates applied to logical qubits. 
(b) Visualizations of the attention matrix at various stages in the circuit, with yellow indicating high attention weight. 
Initially, each qubit attends primarily to itself. After each layer of entangling gates, the attention mechanism dynamically tracks the development of correlations between qubits involved in each gate, as indicated by increased off-diagonal elements. This demonstrates that the attention mechanism learns to focus on pathways along which errors can propagate via entangling operations, providing physical interpretability of the decoder’s decision process.}
\label{fig:interpretability}
\end{center}
\end{figure}
\section{Interpretability}
\vspace{-4mm}

Machine learning decoders are often described as ``black boxes'' due to their complex, opaque internal representations. However, our attention-based neural decoder allows us to probe the model and gain meaningful insight into the error correlations it learns to track. To demonstrate this, we examine the attention weights throughout the layers of an entangling gate circuit on logical qubits, as depicted in \cref{fig:interpretability}(a). After each layer of entangling operations, we extract and visualize the attention weights in the form of an adjacency matrix, which shows the set of qubits that each qubit is most strongly correlated with according to the model’s attention scores (see~\cref{fig:interpretability}(b)). This analysis shows that the decoder dynamically tracks error propagation channels, offering interpretability of the decision process. 

These observations highlight that neural decoders can provide meaningful internal signals aligned with physical intuition. At the same time we note that current analyses capture only coarse correlation structures, leaving open the question of how to systematically extract deeper understanding of the model’s reasoning. Developing frameworks for principled interpretability will be important future work, especially as decoders are deployed in large scale experimental settings.

\section{Discussion and Outlook}
\vspace{-4mm}

We have demonstrated that machine learning decoders constitute a powerful approach for interpreting complex error patterns in quantum circuits. Their practicality depends critically on generating representative training data. For Clifford circuits, such data can be efficiently sampled via classical simulation. The challenge becomes more acute with the inclusion of non-Clifford gates since full quantum state simulation can become intractable. Importantly, our framework circumvents this bottleneck by focusing solely on error propagation and resultant detector outcomes, rather than simulating full quantum states. By carefully tracking how errors propagate through non-Clifford gates, we can continue to efficiently generate necessary syndrome data. This capability substantially broadens the applicability of machine learning-based decoding, bringing even classically non-simulable circuits within reach. Additionally, the interpretability offered by attention mechanisms represents a promising avenue for building understanding in ML decoders. As demonstrated in our analysis, attention weights reveal physically meaningful error correlation patterns that align with expected propagation pathways, providing insight into the decoder's decision-making process and enabling validation of its learned representations.

Despite these advances, scalability remains an open %
problem. As code distance increases, the number of training samples required for reliable performance has been found to grow exponentially~\cite{bausch2024learning}, making training at large distances prohibitively expensive. Addressing this will require the development of more sample-efficient training strategies, such as transfer learning, data augmentation, or architectural innovations aimed at enhancing generalization. We also note that 
our timing analysis (\cref{fig:timing}) shows that general attention-based decoders can achieve $\sim 100\mu$s inference per gate when batched, while more specialized architectures can reach sub-microsecond speeds. Furthermore, these times can be improved through deployment on dedicated hardware such as FPGAs, making real-time decoding within reach for various quantum computing platforms.

Another important direction is the extension to more general classes of codes. Although our current work focused on surface, color, and Reed-Muller codes, the modular design of our neural decoders --- notably the use of graph neural networks for modeling code connectivity --- readily accommodates more complex codes, including topological and low-density parity-check (qLDPC) codes~\cite{breuckmann2021quantum,xu2024constant,xu2025fast,bonillaataides2025constant}. While most progress in qLDPC decoding has focused on simpler memory tasks and noise models~\cite{liu2019neural,maan2024machine,ninkovic2024decoding,gong2024graph,hu2025efficient,blue2025machine}, extending fast, circuit-level decoding to these architectures represents an exciting avenue for future work, requiring new solutions for syndrome processing and error correlation tracking in more intricate code geometries.

Looking ahead, as experiments move toward increasingly complex logical circuits and more realistic noise regimes, the need for fast, accurate, and adaptable decoders will become ever more pronounced. With demonstrated performance on algorithmic circuits and universal gate sets our neural decoder framework offers a strong foundation for this next stage of fault-tolerant quantum computation.

\section{Acknowledgments} 

We acknowledge helpful discussions with Dolev Bluvstein, Sasha Geim, Madelyn Cain, Dom Kufel, Nishad Maskara, Gefen Baranes, Varun Menon, and Nazli Ugur Koyluoglu. We particularly thank Hengyun (Harry) Zhou for providing detailed feedback on our manuscript and results. A.G. thanks Tejasvi Kothapalli for introducing him to the \texttt{schedulefree} optimizer~\cite{defazio2024roadscheduled} used in training the convolutional models for decoding 2D cluster-state circuits. J.P.B.A. and A.G. acknowledge support from the Unitary Foundation. A.G. acknowledges support from the IBM fellowship. We acknowledge support from the National Science Foundation  (PHY-2012023 and CCF-2313084) and  through  the CUA Physics Frontiers Center (PHY-2317134) and NVQL (PHY-2410716) and from DOE through the QUACQ collaboration (DE-SC0025572), Quantum Systems Accelerator Center (DE-AC02-05CH11231), IARPA and the Army Research Office, under the Entangled Logical Qubits program (W911NF-23-2-0219), the DARPA MeasQuIT program (HR0011-24-9-0359).

% \bibliographystyle{apsrev4-2}
% \bibliography{ref}

%apsrev4-2.bst 2019-01-14 (MD) hand-edited version of apsrev4-1.bst
%Control: key (0)
%Control: author (72) initials jnrlst
%Control: editor formatted (1) identically to author
%Control: production of article title (-1) disabled
%Control: page (0) single
%Control: year (1) truncated
%Control: production of eprint (0) enabled
%

\clearpage
\newpage

\section*{Methods}

\noindent\textbf{Architecture}\\

The neural decoder implements a multi-stage architecture that processes syndrome measurements through three conceptually distinct phases: embedding, spatial-temporal processing, and prediction. Unlike traditional decoders that operate on fixed syndrome patterns, our architecture dynamically tracks error correlations as they evolve through the quantum circuit, adapting its internal representations based on both the error correction code structure and the quantum algorithm being executed.

\textit{Core processing pipeline.} The decoder maintains evolving vector representations for each stabilizer measurement location. These representations are updated through alternating spatial and temporal processing layers: spatial layers aggregate information across the code's topology (using convolutions for grid codes like surface codes, or graph neural networks for irregular topologies like color codes), while temporal layers track how errors propagate through circuit layers using gated recurrent mechanisms. This alternating structure allows the model to learn complex spatio-temporal error patterns that arise from coherent errors and correlated noise.

\textit{Input representation and preprocessing.} Raw syndrome measurements are preprocessed through code-specific transformers that preserve topological structure. For surface codes, measurements are arranged on a 2D grid; for color codes, the triangular lattice structure is maintained; for quantum Reed-Muller (RM) codes, specialized graph representations capture the code's connectivity. The preprocessor also constructs dynamic connectivity graphs that change at each time step to reflect transversal gates --- when an entangling gate connects two logical qubits, edges are added between their respective syndrome embeddings, allowing error correlations to propagate.

\textit{Embedding system.} Each stabilizer measurement is embedded into a high-dimensional vector space (for the decoders in this work, the dimension is $384$) by combining multiple information sources: the binary detector value, spatial position within the code, the type of gate applied, noise parameters, temporal position, and (optionally) loss information for erasure events. These diverse features are combined through learnable weighted sums and projected to form the initial stabilizer representations.

\textit{Logical qubit context.} A key innovation is the logical qubit embedding system that tracks correlations between logical qubits. Each logical qubit maintains a ``fingerprint'' embedding that evolves based on the gates applied to it. When entangling gates create correlations between logical qubits, these embeddings are updated through graph neural networks operating on the algorithm's connectivity graph, followed by attention mechanisms and bidirectional recurrent layers. These logical-level embeddings are integrated into the spatial processing, allowing the decoder to distinguish between local physical errors and correlated logical errors.

\textit{Spatial-temporal processing.} The core of the decoder alternates between spatial and temporal updates. Spatial processing uses a combination of local operations (convolutions or graph neural networks depending on code topology) and global attention across all stabilizers. The attention mechanism can dynamically focus on relevant error patterns, with logical qubit embeddings modulating which correlations to track. Temporal processing uses gated updates similar to GRUs but optimized for the discrete time steps of syndrome extraction rounds. Each processing block includes residual connections and normalization for training stability.

\textit{Measurement-basis-aware readout.} The readout module implements sophisticated prediction heads tailored to the measurement basis. It employs dual attention mechanisms with separate parameter sets for X-basis and Z-basis measurements, together with learnable query tokens assigned to each logical observable. The pooled representation is subsequently processed by a multi-layer classifier composed of successive linear transformations interleaved with GELU activations, culminating in a scalar prediction for each logical error probability.

The aggregate parameter count for each of these components together is $13.6$ million parameters.

\noindent\textbf{Training Methodology}\\

\textit{Loss definition.} The job of the neural network decoder is to predict logical flips that have occurred along the readout basis for each logical qubit. The value of this logical flip can be derived by tracking Pauli errors that occur at each step of the circuit, extracting the cumulative Pauli error right before the transversal readout, and calculating the relevant parity. For instance, if the Pauli error afflicting the 3 qubits along the logical observable for a $d=3$ surface code were $ZYI$, and the readout basis was $\bar{X}$, the logical flip would be $0$ (since the number of $Z$-type errors in this error configuration is $2$). Crucially, this method of calculating the logical flip can be used \emph{without reference to the actual value of the measurement outcome}. For algorithms such as simple memories, this is completely equivalent to checking the parity of the final transversal measurement values along the appropriate logical observable. However, for circuits where the transversal measurement value is nondeterministic (even in the absence of noise) --- which is the case for most circuits with magic-state inputs --- the transversal measurement values cannot be used to derive the logical flip, and one must resort to calculating the logical flip using the relevant Pauli errors. With this definition of logical flip, the decoder can be treated as a simple binary classifier, and the training loss is defined simply as a binary cross entropy loss, using the derived logical flip as the true label.

\textit{Data generation pipeline.} Training data is generated online through a simulation pipeline. Random quantum algorithms are sampled with uniform distributions over gate types, circuit depths, numbers of logical qubits, and code parameters. A quantum circuit simulator (Stim) generates syndrome measurements under circuit-level noise, with noise parameters sampled from continuous distributions. The simulation includes realistic effects such as gate errors, measurement errors, and optional qubit loss.

\textit{Optimization and learning rate scheduling.} We employ AdamW optimization with $\beta_1=0.9$, $\beta_2=0.999$, and $\epsilon=10^{-8}$, without weight decay. The learning rate follows a cosine annealing schedule with 1000-step linear warmup:
\begin{equation}
\eta(t) = \begin{cases}
\eta_0 \cdot t/T_{0} & t < T_{0} \\
\eta_{min} + \frac{\eta_0 - \eta_{min}}{2}(1 + \cos(\frac{\pi \cdot (t - T_{0})}{T_{max} - T_{0}})) & t \geq T_{0}
\end{cases}
\end{equation}
where $\eta_0 = 10^{-4}$, $\eta_{min} = 10^{-5}$, $T_{0} = 1000$, and $T_{max} = 10^6$ steps.

\textit{Training strategies for improved generalization.} A key training trick is layered supervision: for a circuit with $n$ layers, the simulator constructs $n$ separate simulation snapshots. The $i$th snapshot contains syndrome measurements for the cumulative circuit up to layer $i$ followed by transversal readout. The decoder processes syndrome measurements from all $n$ snapshots and produces predictions at each layer, with the loss computed across all predictions. This provides $n$ times more training signal per circuit compared to end-to-end supervision, dramatically improving data efficiency and helping the model learn how errors accumulate through circuit layers.  A similar trick is used by ~\citet{bausch2024learning}. Mixed-precision training with half-precision accelerates training while maintaining numerical stability.

\begin{figure}[h!]
    \centering
    \includegraphics[width=\columnwidth]{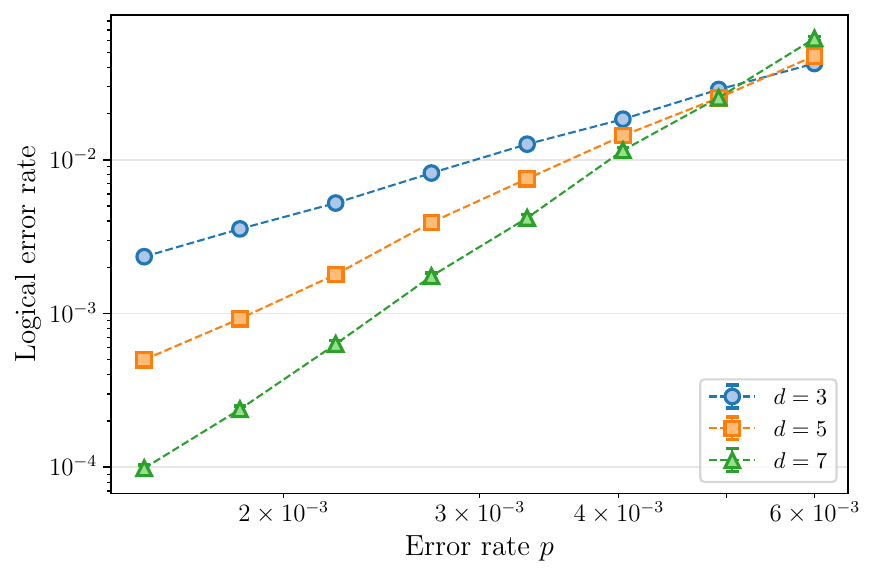}
    \caption{\textbf{Memory performance of 2D color code decoder.} Logical error rate as a function of physical error rate $p$ for 2D color codes of distance $d=3,5,7$, decoded using a graph neural network architecture. The observed threshold of approximately $0.5\%$ matches state-of-the-art results for circuit-level noise. Data are presented as mean values plus or minus standard deviations.}
    \label{fig:color-mem}
\end{figure}

In total, we trained five distinct models corresponding to the figures presented in the main text. The general attention-based decoder for surface codes was used in \cref{fig:neural-decoder}, and was trained with a batch size of 128 for 450000 steps. The loss-aware decoder for the $[[15,1,3]]$ Reed–Muller code (\cref{fig:teleported-memory}) was trained with a batch size of 192 for 200000 steps. For cluster-state circuits (\cref{fig:deep}), both the correlated an uncorrelated decoders with a batch size of 256 for 100000 steps. Finally, a graph neural network decoder was trained for 2D color codes with flag-based extraction (\cref{fig:color-mem}), using a batch size of 256 for 250000 steps. Each model was trained independently with identical optimization settings, differing only in the circuit families and code parameters used for data generation.

\noindent\textbf{Runtime Analysis}\\ We profile inference time to assess scalability. Fig.~\ref{fig:timing} shows inference time for the neural decoder on random Clifford algorithms. Batch-mode inference on NVIDIA H100 GPUs reveals that while theoretical complexity is $O(d^4)$ due to all-to-all attention, empirical scaling is approximately $O(d^2)$ due to GPU optimization. Single-sample inference shows minimal distance dependence due to parallelism. For specialized architectures like the cluster state decoder (Fig.~\ref{fig:deep}), replacing attention with fixed-kernel convolutions yields sub-microsecond inference per qubit, with total decoding time nearly independent of logical qubit count. We note that the single sample inference times can be significantly improved, and brought closer to the amortized runtimes, by deploying the models on dedicated hardware, such as FPGAs.

\noindent\textbf{Other codes}\\
The general algorithmic decoder framework can be extended to a wide range of quantum codes by adapting the spatial update layer (see Fig.~\ref{fig:neural-decoder}(a)). Convolutions are naturally suited for surface codes and other square-lattice variants, given their grid-based spatial structure. For other codes, such as topological color codes, applying standard CNNs is less suitable. For example, forcing a triangular lattice to fit a square grid would necessitate extra padding, complicating the update steps.

A natural and effective generalization is to use graph neural networks (GNNs). By constructing a stabilizer connectivity graph --- where stabilizers are linked if their support overlaps --- the GNN can perform updates analogous to convolutions, but directly on the code's true geometry.

We apply a GNN-based decoder to the 2D color code family, using flag-based syndrome extraction as in Ref.~\cite{Chamberland_2020}. As shown in Fig.~\ref{fig:color-mem}, the decoder achieves a memory circuit threshold near $0.5\%$, on par with state-of-the-art circuit-noise decoding for color codes~\cite{gidney2023new,koutsioumpas2025colour}.

This stabilizer-graph-driven spatial update approach via GNNs is highly versatile---it can be extended to other classes of codes, including high-rate quantum LDPC codes~\cite{bravyi2024high}. A comprehensive investigation of this generalized strategy for broader code families is left for future work.

\noindent\textbf{Numerical simulation details}\\
We now provide additional details regarding the numerical simulations described in the main text.

Fig.~\ref{fig:neural-decoder}(b) shows results for the general algorithm decoder applied to random transversal Clifford circuits on four surface-code logical qubits. The qubits are first initialized in the logical $\ket{\overline{+}}$ state using a noiseless unitary circuit. Subsequent circuit layers feature noisy transversal single-qubit Clifford gates, randomly drawn from the set ${I, X, Y, Z, H}$ on each qubit. Since the Hadamard gate is not transversal in the surface code---it swaps the logical $\overline{X}$ and $\overline{Z}$ operators---we implement it by applying Hadamard gates to all physical qubits, followed by a noiseless lattice rotation to align the logical operators accordingly.

Each layer is followed by a round of two-qubit CNOT gates between randomly selected pairs of logical qubits, applied in random orientations. Although our model architecture includes support for SWAP gate modules, the reported results are restricted to circuits with conventional CNOT gates. After the entangling gate layer, we perform syndrome extraction by measuring the stabilizer checks of the surface code using the standard fault-tolerant pattern.

After the initial noiseless preparation, noise is applied to all physical qubits and gates according to a circuit-level depolarizing noise model. In particular, if we let $p$ characterize the noise strength of the two-qubit gate, so that each of the fifteen errors $\{ IX, IY, \dots , ZZ \}$ occur with the same probability $p/15$, then single qubit gates fail with probability $p/10$, so that each of the three Pauli errors occur with one-third of that probability, and both resets and measurements fail with probability $p/10$. The results in Fig.~\ref{fig:neural-decoder}(b) correspond to a fixed physical error rate of $p=0.3\%$.

For inference, we test decoder performance on circuits of varying depth. For each circuit depth, we sample 20 different random Clifford circuits with the protocol described above. For each sampled circuit, we generate multiple noise realizations (shots) and apply both the trained machine learning (ML) decoder and an optimal most-likely-error (MLE) decoder, which is implemented with gurobi~\cite{cain2024correlated}.

To evaluate decoding accuracy, we focus on logical observables associated with the stabilizer state generated by the circuit. Since simulation of $Y$-basis measurement in the surface code is more complex, our analysis focuses on observables composed solely of $I$, $X$, or $Z$ logical operators. While the ML decoder outputs logical flip probabilities for all logical qubits, the comparison with the MLE decoder is performed by evaluating the parity of a specified target stabilizer after decoding.

The logical error rates presented as a function of circuit depth reflect the fraction of shots on which the decoded parity differed from the true value for the target observable. Reported logical error rates are normalized on a per-round basis.

Fig.~\ref{fig:teleported-memory}(b) shows decoding performance for a logical algorithm in which a qubit encoded in the [[15,1,3]] Reed-Muller (RM) code undergoes repeated layers of transversal single-qubit Clifford gates, Steane-style error correction, and logical teleportation. Ancilla blocks used for syndrome extraction and teleportation are also initialized as [[15,1,3]] RM codes to ensure consistency in error protection throughout the circuit.

During training, we sample circuits of randomly chosen depths, where each layer of single-qubit Clifford gates is selected uniformly from the subset of gates transversal on the RM code: ${ I, X, Y, Z, S }$. We prepare the RM state noiselessly using a unitary preparation circuit. All subsequent operations of the logical algorithm are noisy. The error model assigns the total error budget for each operation between Pauli errors and loss errors, according to the ``loss noise fraction'' parameter shown in Fig.~\ref{fig:teleported-memory}.

Because the RM code is comparatively small, we model its structure with a dense neural network architecture, embedding code-specific features directly into the decoder input. This same approach generalizes to other small quantum codes such as the [[16,6,4]] Tesseract code~\cite{bluvstein2025architectural}.

\end{document}